# No Reduction of Tropical Convection with Warming Expected from Theory or Models

Zachary McGraw[1] and Blaž Gasparini[1]

*1 Department of Meteorology and Geophysics, University of Vienna, Austria*

*Corresponding author: Zachary McGraw, zachary.mcgraw@univie.ac.at*

## Key Points

- Simulations show no clear response of column-integrated convective mass flux to warming, and by level mainly reflect vertical redistribution
- Weakening along isotherms simply reflects reduced mass as isotherms rise into thinner air, whereas convective motions are highly invariant
- A popular hydrological constraint inaccurately quantifies convective weakening at one altitude that is counteracted by strengthening aloft

**Abstract**

Theoretical arguments have for decades anticipated substantial weakening of tropical convection as the surface warms, yet convection-resolving models show no robust change in column-integrated convective mass transport and an increase in transported volume. Here we reconcile this apparent discrepancy by demonstrating that the two main theoretical arguments identify a redistribution, rather than a straightforward reduction, of convective motions. A commonly invoked hydrological argument accordingly captures a weakening at a single altitude that is offset by strengthening aloft as the troposphere expands upward. Further, reduced convective mass flux following atmospheric structures that rise with warming simply reflects decreasing air density aloft. Convective motions and transported water are shown to be highly invariant to surface warming along isotherms, and the ability to predict convective responses at individual altitudes is a manifestation of this invariance. Together, these findings recast and simplify long-standing expectations of how the tropical atmosphere will respond to warming.

**Plain Language Summary**

Convection is the rising motion of warm, moist air that produces clouds and rainfall and helps drive the global circulation. For decades, theories have suggested that convection should substantially weaken across the tropics as the climate warms, raising concerns about how warming will alter tropical rainfall and atmospheric circulation. Yet high-resolution simulations show remarkably little change in the overall amount of convective mass transport. We show that this apparent disagreement largely arises because warming redistributes convection upward. In the thinner atmosphere above, motions transport less mass, giving the appearance of weaker convection even though the motions themselves and water transport change little. A commonly invoked water cycle argument for convective weakening captures a reduction at one altitude yet this is offset by strengthening aloft. These findings suggest that tropical convection is more resilient to warming than previously understood, and redefine expectations for how the tropical atmosphere will change in a warming climate.

## 1. Introduction

Convective mass flux ($M_c$) is a fundamental quantity in tropical climatology that links the collective action of deep convection to the hydrological cycle and the large-scale overturning circulations that shape tropical winds (Held and Soden, 2006; Vecchi and Soden, 2007; Chou and Chen, 2010; Bony et al., 2016). Consequently, the response of $M_c$ to warming is of considerable interest both for anticipating future changes in tropical weather and for advancing understanding of the dynamics governing Earth's climate system.

Theoretical arguments have suggested that convective mass flux should weaken substantially with surface warming, motivating numerous investigations into its temperature sensitivity. One line of reasoning predicts weakening mass flux as the troposphere expands upward into increasingly stable environments, with this weakening identified as robust across models (Jeevanjee, 2022; Williams and Jeevanjee, 2025) and argued to reduce tropical circulation strength (Chou and Chen, 2010). Another argument follows from hydrological constraints: because precipitation is expected to have a weaker fractional response to surface warming than atmospheric humidity, convective mass flux is expected to weaken to maintain the balance between upward water vapor transport and sedimenting condensate. (Held and Soden, 2006; Vecchi and Soden, 2007; Jenney et al., 2020; Jeevanjee, 2022; Williams and Jeevanjee, 2025; Bolot et al., 2025; Mischell, 2026).

Here we show across radiative-convective equilibrium (RCE) simulations that convective mass flux exhibits no straightforward reduction with surface warming. Instead, convective motions are primarily redistributed vertically as the troposphere rises, and neither of the two dominant theoretical arguments for weakened convection expresses more than this result.

## 2. Methods

### *2.1. RCEMIP small domain simulations*

To assess the temperature sensitivity of $M_c$ and its physical controls, we analyze radiative-convective equilibrium (RCE) simulations from the Radiative-Convective Equilibrium Model Intercomparison Project (RCEMIP; Wing et al., 2018). RCE provides an idealized framework for studying convective responses to warming by isolating interactions among convection, radiation, and surface temperature in the absence of processes such as rotation and surface heterogeneity. The limited-area domains used in RCE enable computationally efficient convection-resolving simulations across multiple models, allowing convective mass flux to be directly diagnosed rather than inferred from parameterized convection schemes.

Our analysis builds on the framework of Williams and Jeevanjee (2025; hereafter WJ25), who examined the temperature sensitivity of convective mass flux across the RCEMIP Phase 1 (RCEMIP-I) ensemble. To allow direct comparison to prior findings, we adopt their model selection and diagnostic framework. As in WJ25, we use the RCEMIP-I small-domain simulations, consisting of approximately 100 km × 100 km domains with 1 km horizontal grid spacing and 74 vertical levels extending to 33 km. We use the same models as WJ25 except for MESONH, which we omit due to known humidity diagnostic issues in the RCEMIP-I archive (Mischell, 2026), yielding seven models with the needed output. WJ25 demonstrated that RCEMIP-I exhibits qualitatively similar $M_c$ responses to warming as global climate models, which we do not examine here.

Convective mass flux is calculated following WJ25 by identifying convective updrafts as grid points satisfying both a cloud condensate threshold ($>10^{-5}$ kg/kg) and an upward velocity threshold ($w>1$ m/s). Domain-mean $M_c$ is calculated at each level as the product of $w$ and air density ($\rho$) within convective grid cells, summed across each instance and normalized by fractional horizontal area. Profiles are averaged over 100 three-dimensional snapshots spanning 25 simulated days after a 75-day spin-up period. RCEMIP-I includes simulations with uniform surface temperatures ($T_s$) of 295, 300, and 305 K, allowing responses to warming to be examined. We additionally quantify the full vertical transport of water by resolved motions. Upward water fluxes are obtained by multiplying $M_c$ in each identified convective grid cell by the mixing ratios of water vapor and condensate, including precipitating species; corresponding downward fluxes are calculated over the remaining non-convective (*subsidence*) regions, with domain-mean (*net*) fluxes reflecting both directions.

The above-described diagnostics enable assessment of the hydrological constraint on $M_c$ arising from the contrasting warming responses of precipitation and convective water vapor transport. Because the RCEMIP-I archive does not report vertically resolved latent heating or precipitation flux, we perform separate System for Atmospheric Modeling simulations (SAM; Khairoutdinov and Randall, 2003; Gasparini et al., 2025) in an RCEMIP-I-type configuration. As explained in Text S1, these simulations confirm that precipitation flux at every level closely balances the total net upward water flux by resolved motions at kilometer-scale resolution, supporting the use of the available RCEMIP-I diagnostics in our analysis.

## 3. Results

### *3.1. No simulated reduction of convection in response to surface warming*

We begin by examining the response of convective mass flux $M_c$ to surface warming across the RCEMIP-I simulations. While WJ25 primarily assessed the sensitivity of $M_c$ to surface warming along isothermal coordinates, we focus here on the response of $M_c$ in standard height coordinates and its integral across the atmospheric column. Fig. 1a shows $M_c$ profiles from simulations having surface temperatures of 295 K (solid lines) and 305 K (dashed lines), with black lines indicating the ensemble-mean profiles and colored lines showing individual simulations at $T_s$ = 295 K. Fig. 1b shows the corresponding sensitivities of $M_c$ to surface warming, expressed as percent change per degree Kelvin and calculated as 100% × $[(|X_{305K}|/|X_{295K}|)^{1/\Delta T} - 1]$ throughout this study.

Firstly, it is apparent in Fig. 1b that the ensemble-mean response (black line) changes sign across the troposphere, with slight weakening in the low-to-mid troposphere offset by strengthening aloft. Comparing the ensemble-mean $M_c$ profiles at $T_s$ = 295 K and 305 K (black solid and dashed lines in Fig. 1a, respectively) suggests that much of this response simply reflects redistribution of the $M_c$ profile as the troposphere expands upward with warming. Beyond this, the $M_c$ responses are modest: the strongest responses in individual models (colored lines) are not robust, with no level showing a decrease exceeding 1 %/K consistently across models. While WJ25 identified the redistribution as one factor contributing to the $M_c$ response to warming, we demonstrate here and in the next section that it accounts for the large majority of the response.

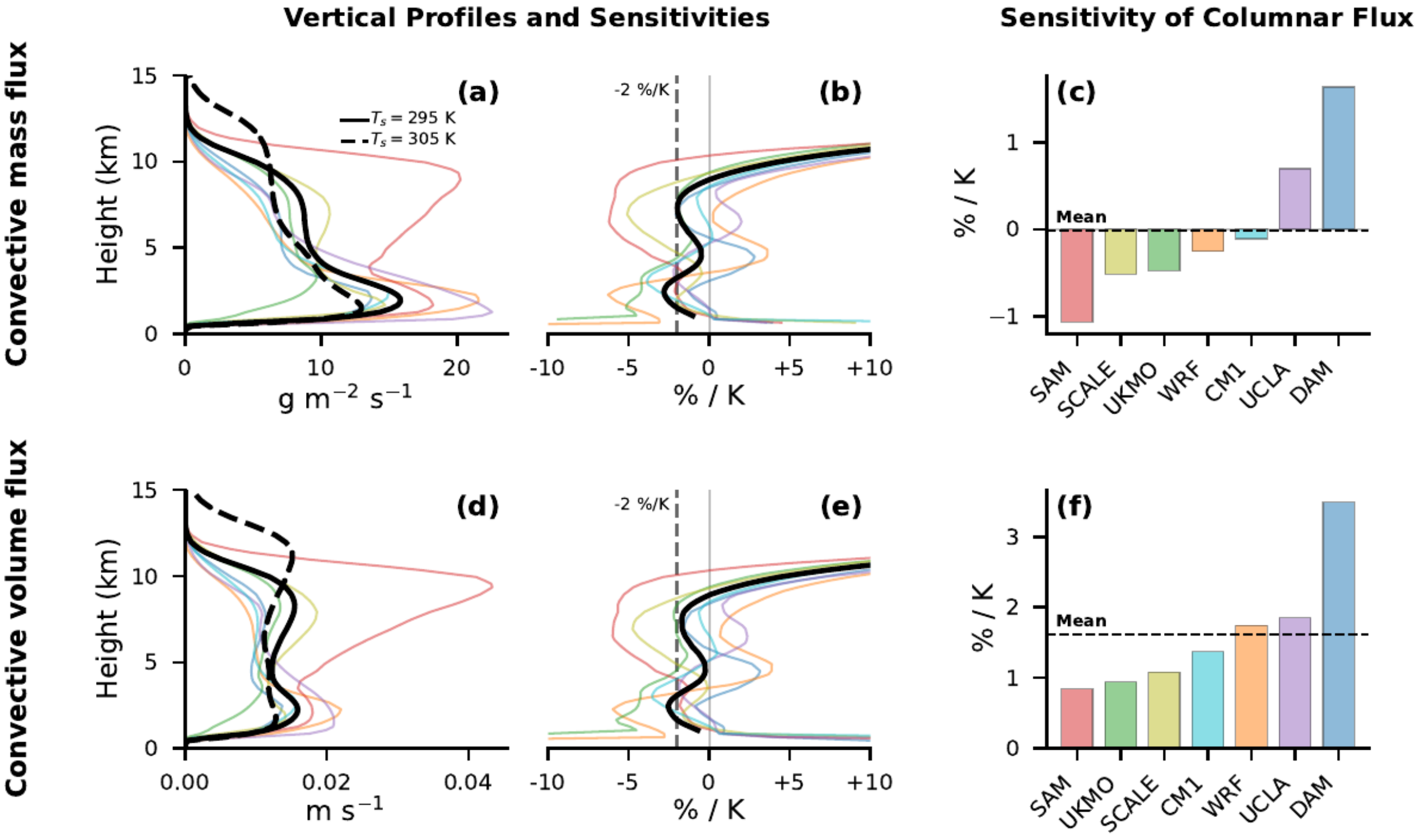

**Figure 1.** Vertical profiles of convective mass flux ($M_c$; top row) and volume flux ($V_c$; bottom row) at 295 K and 305 K surface temperatures, shown as absolute values (a,d), warming sensitivities (% / K; b,e), and column-integrated values (c,f). Black lines indicate ensemble-mean values, while colored lines indicate individual models; individual profiles in (a,d) are shown only for the 295 K simulations. Model colors in (a-d) are consistent with those shown in (c,f). In (b,e), ensemble-mean sensitivities are omitted in the lowest 1 km, where values are noisy due to limited convective activity. Upper-tropospheric sensitivities extend beyond the plotting range because warming produces convection at levels where it is weak or absent in the reference climate.

The lack of a straightforward reduction is more clearly evident when integrating $M_c$ vertically to calculate the total mass transport through the atmospheric column (Fig. 1c), following

Chadwick et al. (2013) but with resolved rather than parameterized convective fluxes. Crucially, the column-integrated response across RCEMIP-I is negligible: the sign of the response is mixed across the seven assessed models, only one decreases as much as 1 %/K, and the mean response across the models is very nearly zero. Thus, it is clear that warming does not clearly reduce the total mass that is transported upward by convection; instead, it acts as a form of *stretching* that redistributes $M_c$ vertically.

Because $M_c$ couples the velocity of convective motions with the density of the transported air, we additionally examine convective volume flux, $V_c = M_c / \rho$ in units of $m^3/m^2/s$ (or equivalently m/s), as a more direct measure of convective motion at each level and overall. While the sensitivity of $V_c$ to warming at each level (Fig. 1e) is similar to that of $M_c$, column-integrated $V_c$ *increases* in all RCEMIP-I models, with an ensemble-mean response of +1.7 %/K that compared to $M_c$ does not capture how warming slightly reduces $\rho$. Hence, notions that surface warming weakens convection hinge on the metric assessed, and when viewed in the upper troposphere or in terms of total volume lifted convection actually *strengthens*.

### *3.2. Reframing the isothermal view: invariant motions, with thinner air aloft*

Having established the absence of a clear convective weakening across simulations, we now reconcile this with the first major argument for why convection should weaken with warming: the upward displacement of convection into an increasingly stable environment (Knutson and Manabe, 1995; Chou and Chen, 2010). This stability-modulated weakening has recently been shown to be robust across models when $M_c$ is evaluated along isotherms (Jeevanjee, 2022; WJ25), with this coordinate reflecting the upward displacement of the free-tropospheric structure with warming such that corresponding levels remain at approximately the same temperature (Hartmann and Larson, 2002; Jeevanjee and Romps, 2018). One might without further context interpret the isothermal reduction as a pattern hidden in the noisy $M_c$ geometric profiles across simulations (Fig. 1a,b). This, we will demonstrate, is not the case.

To reveal the physics underpinning the isothermal $M_c$ response, we show in Figure 2 a key distinction: while the vertical flux of many atmospheric quantities  associated with moist convection remain nearly unchanged along isotherms, $M_c$ exhibits a substantial decline. The ensemble-mean vertical fluxes of water mass, whether vapor (Fig. 2a,f) or condensate (Fig. 2b,g), change by less than 1 %/K across the free troposphere. While volume flux ($V_c$) was shown by Jeevanjee (2022) to weaken along isotherms in one atmospheric model, it exhibits a near-zero ensemble-mean response in RCEMIP-I across the free troposphere (Fig. 2c,h). In contrast, $M_c$ decreases by roughly 3 %/K along isotherms in all models (Fig. 2d,i), as was identified by WJ25.

The explanation is straightforward: *as isotherms move aloft with surface warming, invariant motions move less mass in thinner air*. Since $M_c = V_c\,\rho$, and $V_c$ is isothermally invariant, the $M_c$ response closely mirrors the isothermal reduction in air density (Fig. 2e,i). The apparent weakening of $M_c$ therefore simply reflects that air becomes thinner aloft, rather than a weakening of convective motion. Further, the isothermal invariance of convective motions revealed by $V_c$ is not a peculiarity of this metric: latent heating and vertical energy transport by convection also exhibit isothermal invariance (Sokol and Hartmann, 2022). Together, these simulation results suggest that the collective effect of convective motions is to first-order fundamentally invariant along isotherms.

This reframing highlights the importance of disentangling coordinate effects from physical changes within climate science. Isothermal coordinates naturally expose the invariance of quantities central to moist convection whose behavior is governed primarily by water, including latent heating, cloud distributions, and atmospheric radiative cooling (Hartmann and Larson, 2002; Jeevanjee and Romps, 2018). By contrast, $M_c$ is weighted by dry-air mass, which plays a relatively passive role in regulating moist convective energetics and transport.

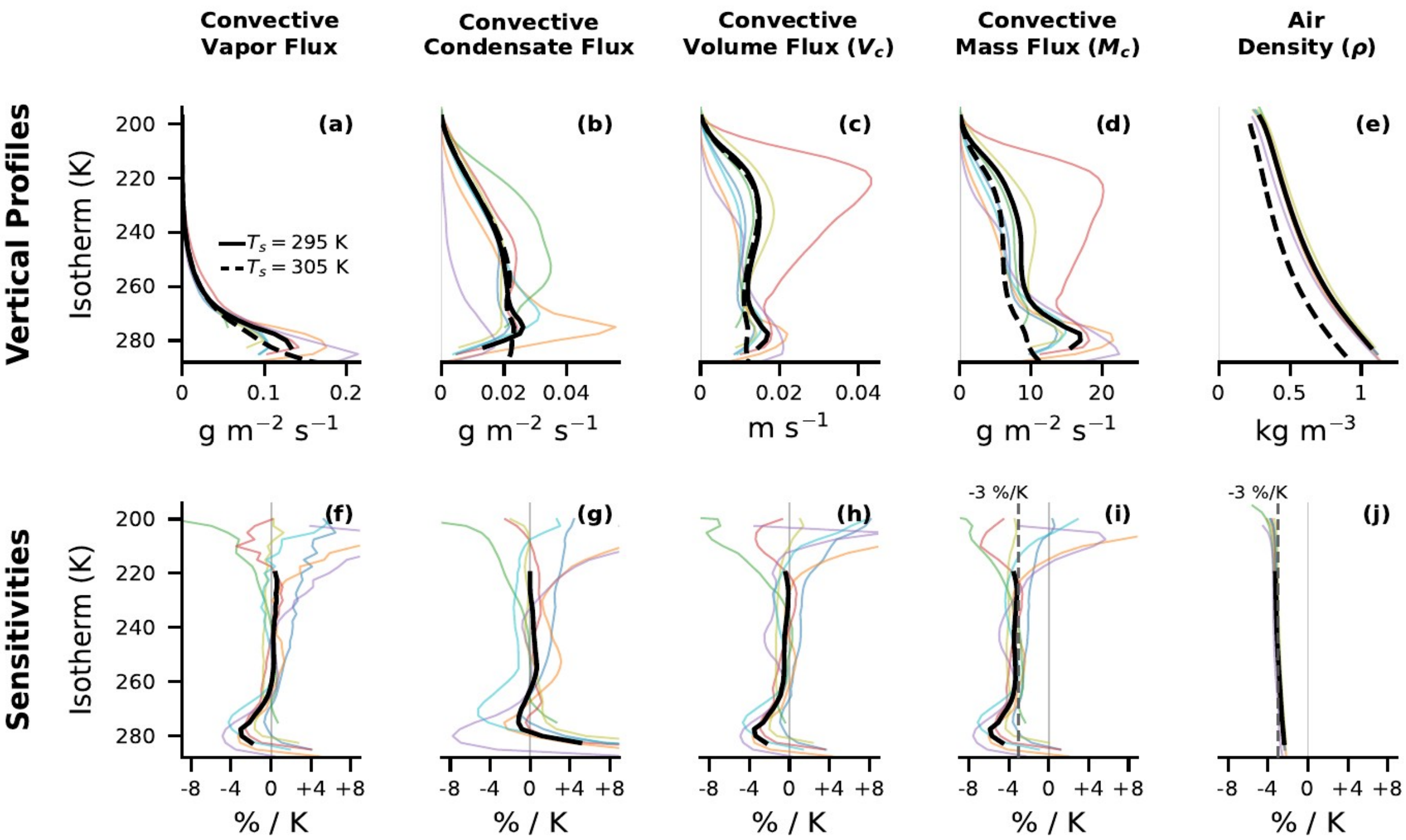

**Figure 2.** Key variables expressed in isothermal coordinates, shown as vertical profiles (a–e) and their sensitivity to surface warming (f–i). In (a–e), solid lines indicate the 295 K simulations and dashed lines the 305 K simulations, with individual model profiles shown for the 295 K ensemble. Black lines indicate ensemble means. In (e–i), data are omitted below the low-level maximum in cloud fraction for each model, while ensemble means are omitted where this masking affects more than 1 model, following WJ25. Ensemble means are also omitted above the 220 K isotherm, where intermodel spread becomes too large for a meaningful ensemble estimate.

In every model, the isothermal reduction in air density increases approximately linearly with height, from ~2 %/K near the surface to ~4 %/K near the tropopause, yielding an approximately 3 %/K reduction when averaged across the troposphere (Fig. 2i). Eqn. 12 of WJ25 predicts the fractional pressure response to warming, which, by the ideal gas law, equals the fractional density response along isotherms. Their framework uses a bulk lapse rate from the surface to the tropopause, yielding a ~2 %/K contribution associated with the magnitude of the bulk lapse rate and an additional term associated with its response to warming. We instead formulate the response locally. The height of an isotherm responds to each infinitesimal surface warming according to the local atmospheric warming at its initial height and the local environmental lapse rate ($\Gamma$):

$$\frac{dz(T,T_s)}{dT_s} = \frac{\left(\partial T/\partial T_s\right)_{z=z(T,T_s)}}{\Gamma(T,T_s)} \qquad \text{Eqn. 1}$$

Applying hydrostatic balance and the ideal gas law then translates this into the fractional air density response per K of surface warming:

$$\frac{1}{\rho}\frac{d\rho(T,T_s)}{dT_s} = -\frac{g}{R_d\, T\, \Gamma(T,T_s)}\left(\frac{\partial T}{\partial T_s}\right)_{z=z(T,T_s)} \qquad \text{Eqn. 2}$$

Eqn. 2 shows that the air density response is the product of two factors: the local atmospheric warming per degree of surface warming, and a ~2 %/K factor set by the local lapse rate. The local warming varies from ~1–2 K/K across the troposphere, such that calculating the product from the 295 K to 305 K simulations replicates the replicating the simulated $\rho$ response (decomposition shown in Fig. S1) that is the main driver of the $M_c$ response.

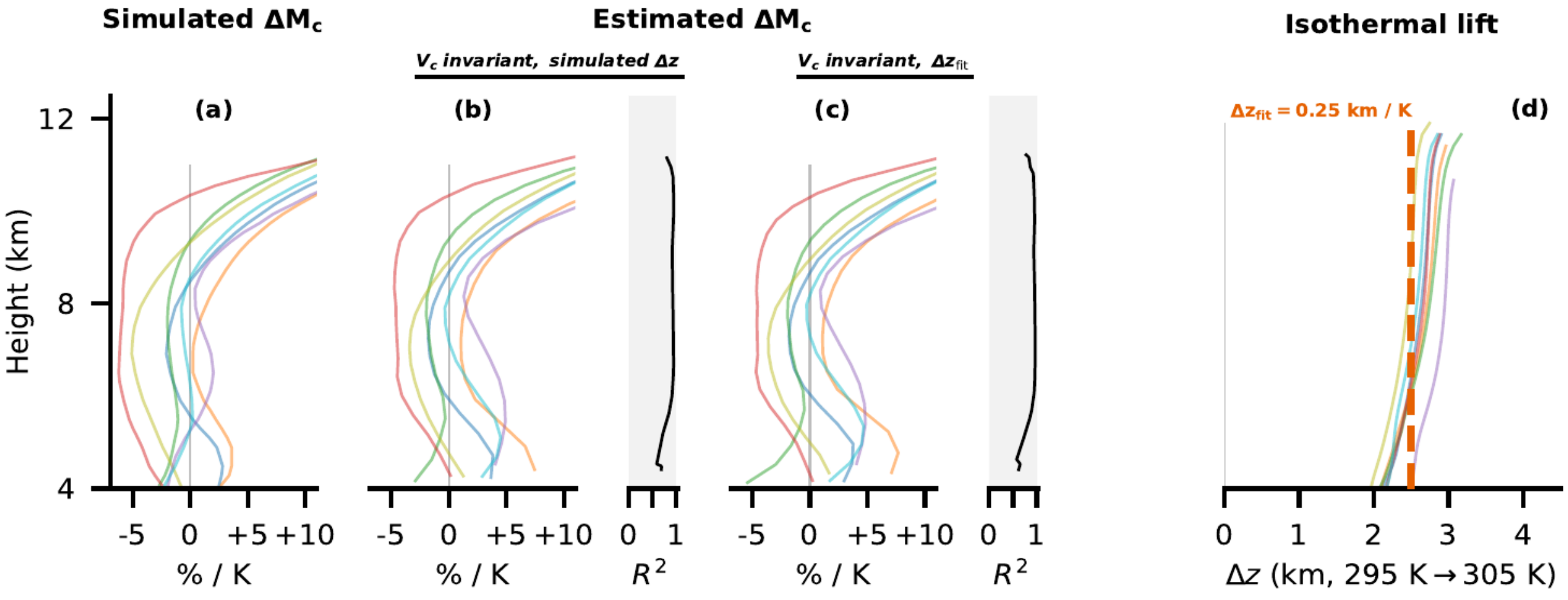

**Figure 3.** Predicted $M_c$ response estimated solely from invariant $V_c$ and the upward displacement of isotherms as surface temperature increases from 295 K to 305 K. Actual simulated $M_c$ sensitivities at each altitude above 4 km are shown in (a). Predictions using the simulated isothermal rise are shown in (b), while (c) uses a crude assumption of 0.25 km/K

that approximately reproduces the simulated isothermal displacement (d). In (b,c), levels are omitted where the isotherm entering a given level at $T_s$ = 305 K is warmer than 280 K, because isothermal invariance is strongest in the free troposphere, motivating the 4 km cutoff.

We now turn to where the isothermal perspective *does* provide value for predicting how $M_c$ changes with warming. WJ25 developed an analytical framework combining thermodynamic constraints, lapse-rate changes, and $M_c$ profile structure, and showed that it can successfully predict the response of $M_c$ at each altitudes across models. Here we demonstrate that this predictability can be viewed more simply: the response is determined almost entirely by the initial-state $V_c$ profile.

Figure 3 shows that the simulated $M_c$ sensitivity across RCEMIP-I can be accurately reproduced by assuming that $V_c$ at the 295 K base state remains invariant along isotherms and accounting only for how far each isotherm rises between the 295 K and 305 K simulations. Intermodel differences in $M_c$ sensitivity are so dominated by differences in $M_c$ profile shape that this approach closely reproduces the simulated responses ($R^2$ > 0.9 across most of the free troposphere) whether inputting the simulated isothermal rise (Fig. 3b) or simply assuming a uniform rise 0.25 km/K (Fig. 3c), which roughly matches the simulated response (Fig. 3d).

Given the above, we reframe the predictability result of WJ25 as follows: rather than reflecting complex and dynamically important convective changes, this skill is a testament to convective *invariance* along isotherms. Convective motions move upward along isotherms, and the slope of the rising $V_c$ profile is the dominant control of how $M_c$ changes at each altitude. Likewise, the reduction in $M_c$ along isotherms has no predictive skill for how $M_c$ changes at any height, as is apparent when separating the thermodynamic and profile shape terms of the WJ25 analytical model (their Eqn 13.), which unveils that all predicted model difference lies in the latter (see Fig. S2).

Our reframing of the isothermal $M_c$ perspective calls into question the physical significance of theorized convective weakening. Jeevanjee (2022) and WJ25 attributed the reduction in $M_c$ along isotherms to lapse-rate stabilization, mirroring the argument of Knutson and Manabe (1995) that $M_c$ weakens as static stability increases with warming and the attribution by Chou and Chen (2010) of this stabilization to tropospheric expansion. The results here suggest a physically simpler picture: convective motions are highly invariant to surface warming along isotherms, while the response at any fixed altitude (or pressure) reflects convective motions in underlying layers being displaced upward as the troposphere expands.

### *3.3. Is a Hydrological Constraint on Convective Mass Flux Insightful?*

Now that we have shown that one of the main predictions of the convective response to warming does not imply a reduction, we turn to the other: that the smaller fractional response of precipitation than atmospheric humidity to warming implies a decrease in convective mass flux. This relationship is typically expressed following Held and Soden (2006; hereafter *HS06*) as

$$P_{sfc} \approx M_{c,b}\, q_{v,b} \qquad \text{Eqn. 3}$$

where surface precipitation ($P_{sfc}$) is balanced by convective water vapor transport at a representative cloud-base level, denoted by the subscript $b$. The transport is approximated as the product of convective mass flux ($M_{c,b}$) and the environmental water vapor mixing ratio ($q_{v,b}$) at that level. Applying this relationship with the ~2-3 %/K precipitation response found across climate models and the ~7 %/K water-vapor increase expected from Clausius-Clapeyron scaling (Allen and Ingram, 2002) implies an approximately 4-5 %/K reduction in $M_{c,b}$. In this section we build on other recent studies using RCEMIP-I models to examine this constraint (WJ25; Mischell, 2026), here reconciling how it anticipates substantial convective weakening when simulations overall lack a reduction.

Firstly, $M_{c,b}$ responses inferred by the HS06 constraint do not imply that convection is inherently reduced with warming. Crucially, the constraint is an attempt to reconcile model results at a single level of the atmosphere. Our results show that warming redistributes a near-constant amount of convective mass flux across an expanding troposphere (Section 3.1), such that the response at any fixed altitude (or pressure) in the free troposphere reflects convective motions in underlying layers being displaced upward to maintain the same temperature (Section 3.2). Because atmospheric layers near the base of the free troposphere themselves rise with warming, even if not fully isothermally, the HS06 scaling primarily captures a redistribution of $M_c$ (Fig. 1a) rather than a reduction. Hence, even if the HS06 constraint were fully accurate, it could only diagnose convective weakening at one level, which is expected to be counteracted by strengthening aloft.

Second, layers that could be interpreted as cloud base undergo the strongest convective weakening by level across the troposphere. Whereas the troposphere-integrated RCEMIP-I $M_c$ response is near zero (Fig. 1c), $M_c$ reductions peak in the lower troposphere. The peak ensemble-mean reduction of -2.8 %/K (black line in Fig. 1b) coincides with the abrupt transition from negligible $M_c$ to the overlying tropospheric $M_c$ maximum near 2 km (black lines in Fig. 1a): as the troposphere expands, $M_c$ is redistributed upward from the lowest convective layers, with little underlying $M_c$ available to replace it. Thus, cloud-base $M_c$ overstates convective weakening by sampling where upward redistribution produces its strongest local signature.

And third, the HS06 constraint is inaccurate, being systematically prone to overpredict the weakening of $M_{c,b}$. Figure 4a shows that in 4 of the 7 assessed models there is not a single altitude at which the simulated $M_c$ response (brown lines) is as negative as that inferred from the ratio (solid green lines) of surface precipitation to local environmental water vapor mixing ratio. Using the local precipitation flux at each level – here set to the balancing net advective water flux due to direct precipitation flux output being unavailable from RCEMIP-I – provides a more appropriate constraint by balancing the factors at the same level, yet produces a similar bias. Because we diagnose the full budget of vertical water transport by resolved motions (Section 2.1), we can identify the source of this bias: the HS06 water budget approximation omits substantial compensating responses. Figure 4b shows the ensemble-mean water vapor flux, with total advected vapor (orange lines) comprising upward transport in convective regions (gold lines) and downward transport in subsidence regions (red lines); Fig. 4c shows the corresponding net transport of vapor (orange lines) and condensate (blue lines), which together constitute the total advected water flux. The key result is that the total advected water that balances precipitation is less sensitive to warming than convective water vapor transport, biasing the results of the HS06 scaling toward excess $M_c$ weakening. This contrast is evident at every level of the troposphere in the ensemble mean (gold versus black in Fig. 4d), and in 5 of the 7 models individually (Fig. S4). We note that subsiding vapor transport was proposed in the earlier Betts (1998) formulation on which HS06 was based, but WJ25 found this more complex constraint less predictable, and our analysis identifies the limited response of condensate to warming as a further complication.

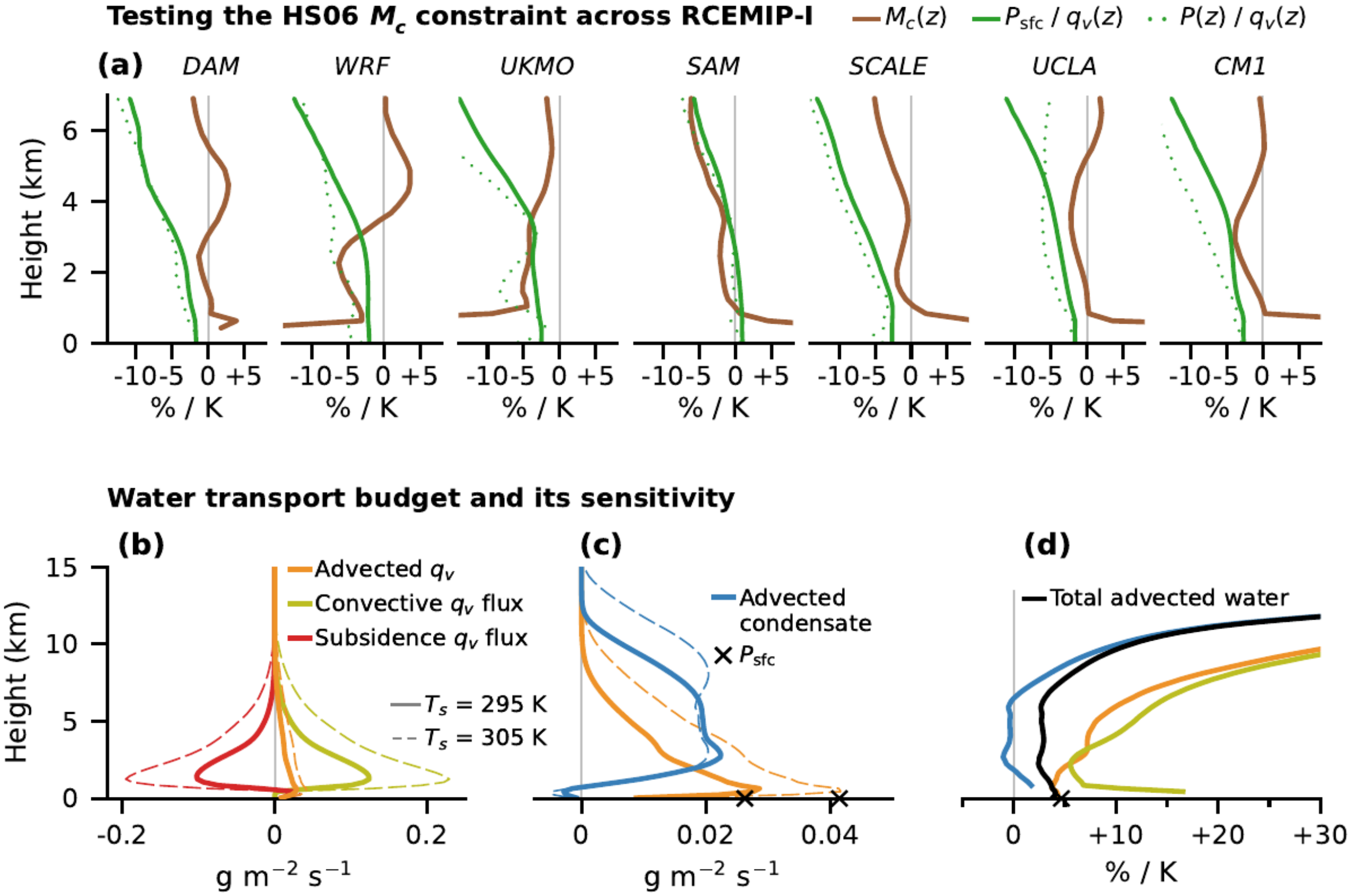

**Figure 4.** Comparison between the simulated response of $M_c$ and that predicted by the HS06 hydrological scaling (a), together with an explanation of bias based on the atmospheric water transport budget (b–d). Panels (b–e) show ensemble-mean profiles and sensitivities averaged across the seven assessed RCEMIP-I models. Black × symbols in (c) and (d) denote the surface precipitation rate $P_{sfc}$, which is the only direct precipitation diagnostic available in the archived RCEMIP-I output.

Hence, while prior studies have identified limitations of the hydrological constraint on $M_c$ (Schneider et al., 2010; Jeevanjee, 2022; WJ25; Mischell, 2026), we show that the limitations are fundamental: its one-level formulation is unrepresentative of the overall convective response and its quantitative predictions are systematically biased due to an oversimplified water cycle representation. Even setting aside the quantitative bias, the HS06 constraint does not appear to imply any weakening of convection that is not counteracted by a strengthening at higher levels.

## 4. Conclusions

This study challenges the prevailing view of how tropical convection responds to surface warming. We have demonstrated that simulations show no straightforward reduction of convective mass flux, while long-established theoretical arguments instead capture a redistribution of convective motions: a commonly invoked hydrological constraint estimates a local reduction that is offset aloft, while convective mass flux weakens along rising atmospheric structures because air density declines aloft while convective motion does not.

How this reframing applies to individual storms remains unknown. Recent work finds that storm intensity and structure respond to warming differently than expected from the bulk tropical atmosphere (Bolot et al., 2025). Upcoming satellite missions, including INCUS (e.g., Dolan et al., 2023), C²OMODO (e.g., Puech et al., 2023), and WIVERN (Illingworth et al., 2018), will provide observations of vertical motions and convective structure, enabling direct estimates of convective mass transport across storms and environments. These measurements could enable tests of whether the convective invariance identified here is realized in nature across the tropics and at smaller scales.

It also remains uncertain what this reframing implies for the large-scale tropical overturning circulation, which balances convective mass transport through divergent horizontal winds. Our results suggest that circulation responses inferred from interpretations of convection may likewise require reframing (Chou and Chen, 2010; Held and Soden, 2006). However, the idealized RCEMIP-I simulations cannot directly assess large-scale circulation responses.

Global storm-resolving models could overcome this limitation and be used to identify what the invariance of convective motions implies for their coupling to the circulation.

## Acknowledgments

The authors thank Aiko Voigt, Andrew IL Williams, and Lorenzo Polvani for useful discussions.

## Open Access Statement

Data analyzed in this study are from the Radiative-Convective Equilibrium Model Intercomparison Project Phase 1 (RCEMIP-I; Wing et al., 2018) archive and are publicly available through the RCEMIP-I data repository. Output from the SAM-P3ice simulations used to test closure of the atmospheric water transport budget is archived at McGraw (2026).

## Conflict of Interest

The authors attest that they have no conflicts of interest.

## References

Allen, M. R., & Ingram, W. J. (2002). Constraints on future changes in climate and the hydrologic cycle. Nature, 419(6903), 224-232.

Bolot, M., Roca, R., Fiolleau, T., & Muller, C. (2025). No decrease of tropical convection in individual deep convective systems with global warming. npj Climate and Atmospheric Science.

Chadwick, R., Boutle, I., & Martin, G. (2013). Spatial patterns of precipitation change in CMIP5: Why the rich do not get richer in the tropics. *Journal of climate*, *26*(11), 3803-3822.

Chou, C., & Chen, C. A. (2010). Depth of convection and the weakening of tropical circulation in global warming. *Journal of Climate*, *23*(11), 3019-3030.

Dagan, G., Koren, I., Altaratz, O., & Feingold, G. (2018). Feedback mechanisms of shallow convective clouds in a warmer climate as demonstrated by changes in buoyancy. Environmental Research Letters, 13(5), 054033.

Dolan, B., Kollias, P., van den Heever, S. C., Rasmussen, K. L., Oue, M., Luke, E., ... & Chandrasekar, V. (2023). Time resolved reflectivity measurements of convective clouds. *Geophysical Research Letters*, *50*(22), e2023GL105723.

Gasparini, B., Atlas, R., Voigt, A., Krämer, M., & Blossey, P. N. (2025). Tropical cirrus evolution in a kilometer-scale model with improved ice microphysics. Atmospheric Chemistry and Physics, 25(17), 9957-9979.

Hartmann, D. L., & Larson, K. (2002). An important constraint on tropical cloud-climate feedback. *Geophysical research letters*, *29*(20), 12-1.

Held, I. M., & Soden, B. J. (2006). Robust responses of the hydrological cycle to global warming. Journal of climate, 19(21), 5686-5699.

Illingworth, A. J., Battaglia, A., Bradford, J., Forsythe, M., Joe, P., Kollias, P., ... & Wolde, M. (2018). WIVERN: A new satellite concept to provide global in-cloud winds, precipitation, and cloud properties. *Bulletin of the American Meteorological Society*, *99*(8), 1669-1687.

Jeevanjee, N., & Romps, D. M. (2018). Mean precipitation change from a deepening troposphere. *Proceedings of the National Academy of Sciences*, *115*(45), 11465-11470.

Jeevanjee, N. (2022). Three rules for the decrease of tropical convection with global warming. *Journal of Advances in Modeling Earth Systems*, *14*(11), e2022MS003285.

Jenney, A. M., Randall, D. A., & Branson, M. D. (2020). Understanding the response of tropical ascent to warming using an energy balance framework. Journal of Advances in Modeling Earth Systems, 12(6), e2020MS002056.

Khairoutdinov, M. F., & Randall, D. A. (2003). Cloud resolving modeling of the ARM summer 1997 IOP: Model formulation, results, uncertainties, and sensitivities. Journal of the Atmospheric Sciences, 60(4), 607-625.

McGraw, Z. (2026). SAM-P3ice water budget closure archive (295 and 305 K) (Version v1) [Data set]. Zenodo. https://doi.org/10.5281/zenodo.21924973

Mischell, E. (2026). Perspectives on Storm Dynamics and Precipitation in the General Circulation of the Atmosphere (Doctoral dissertation, University of Miami).

Morrison, H., & Milbrandt, J. A. (2015). Parameterization of cloud microphysics based on the prediction of bulk ice particle properties. Part I: Scheme description and idealized tests. Journal of the Atmospheric Sciences, 72(1), 287-311.

Puech, J., Hermozo, L., Brogniez, H., Roca, R., Fiolleau, T., Chaboureau, J. P., ... & Krieg, J. M. (2023). The C2Omodo Concept: A Tandem of New Generation High Resolution All-Sky Atmospheric Sounders in the Frame of the AOS Mission. In *IGARSS 2023-2023 IEEE International Geoscience and Remote Sensing Symposium* (pp. 4576-4579). IEEE.

Schneider, T., O'Gorman, P. A., & Levine, X. J. (2010). Water vapor and the dynamics of climate changes. *Reviews of Geophysics*, *48*(3).

Sokol, A. B., & Hartmann, D. L. (2022). Radiative cooling, latent heating, and cloud ice in the tropical upper troposphere. *Journal of climate*, *35*(5), 1643-1654.

Williams, A. I., & Jeevanjee, N. (2025). A robust constraint on the response of convective mass fluxes to warming. Journal of Advances in Modeling Earth Systems, 17(4), e2024MS004695.

Wing, A. A., Reed, K. A., Satoh, M., Stevens, B., Bony, S., & Ohno, T. (2018a). Radiative–convective equilibrium model intercomparison project. *Geoscientific Model Development*, *11*(2), 793-813.

Wing, A. A., Stauffer, C. L., Becker, T., Reed, K. A., Ahn, M.-S., Arnold, N. P., et al. (2018b). *Radiative-Convective Equilibrium Model Intercomparison Project (RCEMIP) Simulation Dataset* [Data set]. World Data Center for Climate (WDCC). https://www.wdc-climate.de/ui/info?site=RCEMIP_DS

***Text S1*** | *Assessing Water Transport Budget Closure with SAM-P3ice Simulations*

To assess whether vertically-resolved water transport budgets can reliably be reconstructed from the diagnostics available in the RCEMIP-I archive, we perform additional RCEMIP-I-style simulations with the System for Atmospheric Modeling (SAM; Khairoutdinov and Randall, 2003). The RCEMIP-I archive does not include precipitation sedimentation fluxes or water transport by parameterized subgrid-scale processes, preventing direct evaluation of budget closure. Our SAM-P3ice simulations provide these additional diagnostics, allowing us to quantify the resulting residual and assess whether resolved advected water transport provides a sufficiently accurate approximation to the total upward water flux.

Relative to the published RCEMIP-I SAM simulations, our simulations use a slightly expanded horizontal domain of 128 km and six additional vertical levels, extending the model top to 36 km. We employ the P3 microphysics scheme (Morrison and Milbrandt, 2015), including recent ice-microphysics modifications described by Gasparini et al. (2025), rather than the simpler one-moment microphysics used in the RCEMIP-I SAM simulations. Throughout this study, “SAM” refers exclusively to the original RCEMIP-I SAM simulations; results from SAM-P3ice simulations are reported only in this text and Fig. S3.

In SAM-P3ice, precipitation flux closely matches the net upward transport of non-precipitating water by resolved motions at all levels (purple and black lines in Fig. S3). The main exception is the lowest atmospheric layer, where parameterized transport dominates (green line). However, because surface precipitation is available for all RCEMIP-I models, the absence of parameterized subgrid-scale boundary-layer transport is unlikely to substantially affect our analysis. Thus, resolved advective water transport captures nearly all of the upward water flux balanced by precipitation, supporting the use of the advected water fluxes available in RCEMIP-I (Section 2.1) to examine the physical basis of the hydrological constraint on convective mass flux proposed by Held and Soden (2006).

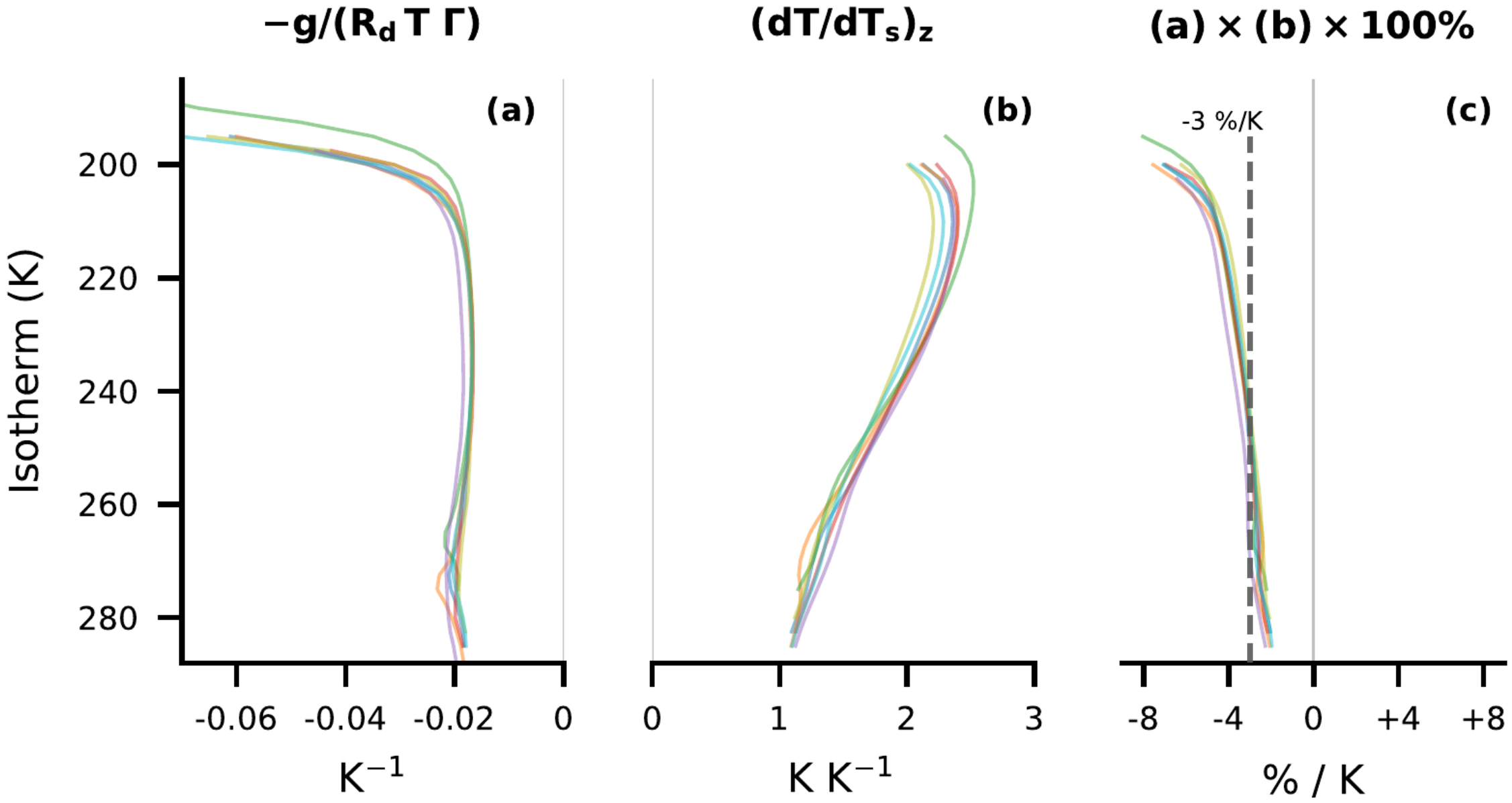

**Fig. S1** | Local drivers of the reduction in air density along isotherms, following Eqn. 2. Panel (a) is derived from the lapse rates at $T_s$ = 295 K, while panel (b) uses the temperature difference from 295 K to 305 K. Each colored curve shows values from an individual model, following the colors used for each model in Fig. 1c.

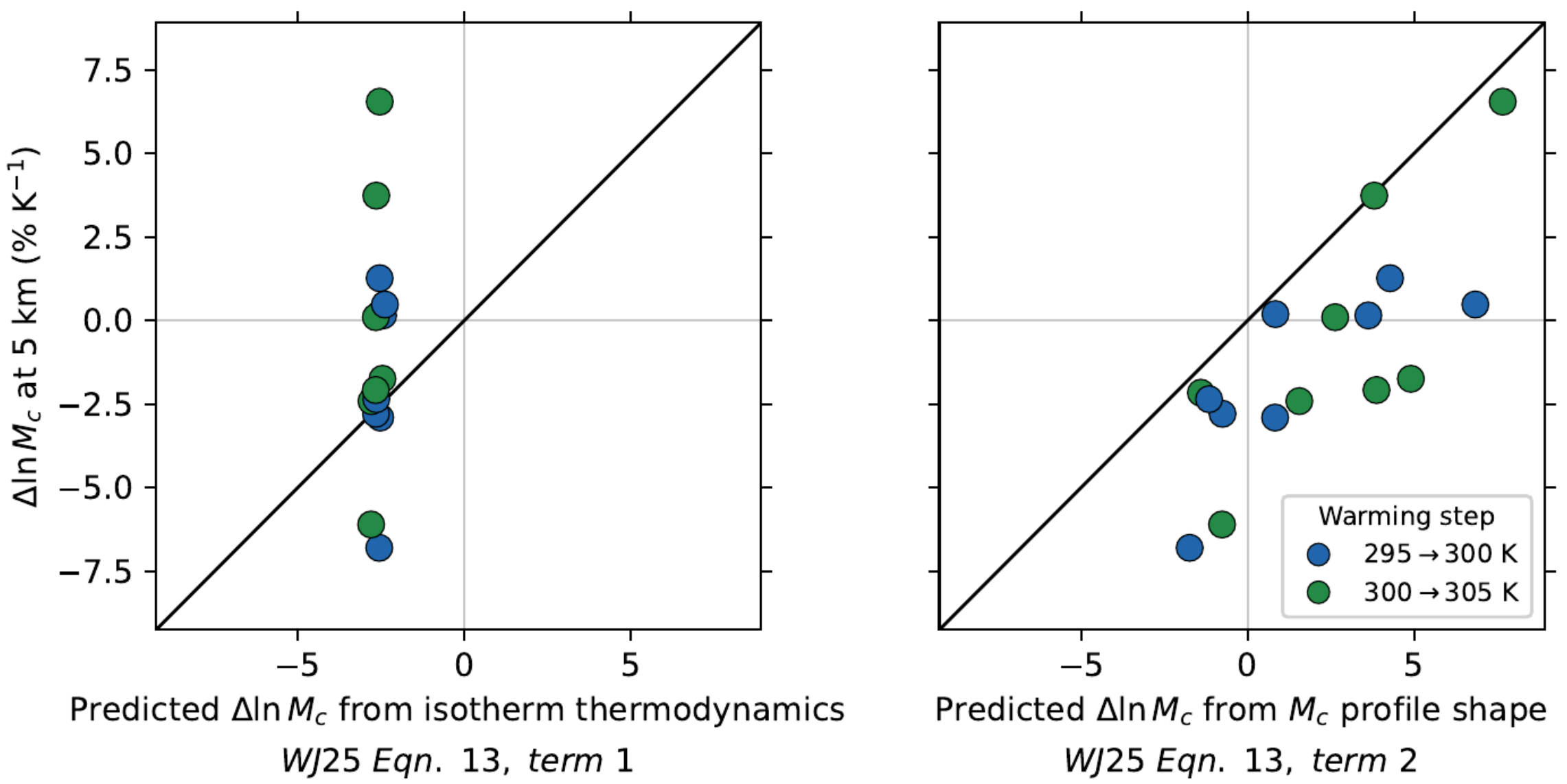

**Fig. S2 |** Prediction of percent-per-Kelvin response of $M_c$ ($\Delta\ ln\ M_c$) following Eqn. 13 of WJ25 and their Fig. 5, but split in terms to show that all predicted intermodel spread is linked to Mc profile shape (term 2, at right). Note that to stay faithful to WJ25, we here include MESONH model output from RCEMIP, which we exclude otherwise in this study.

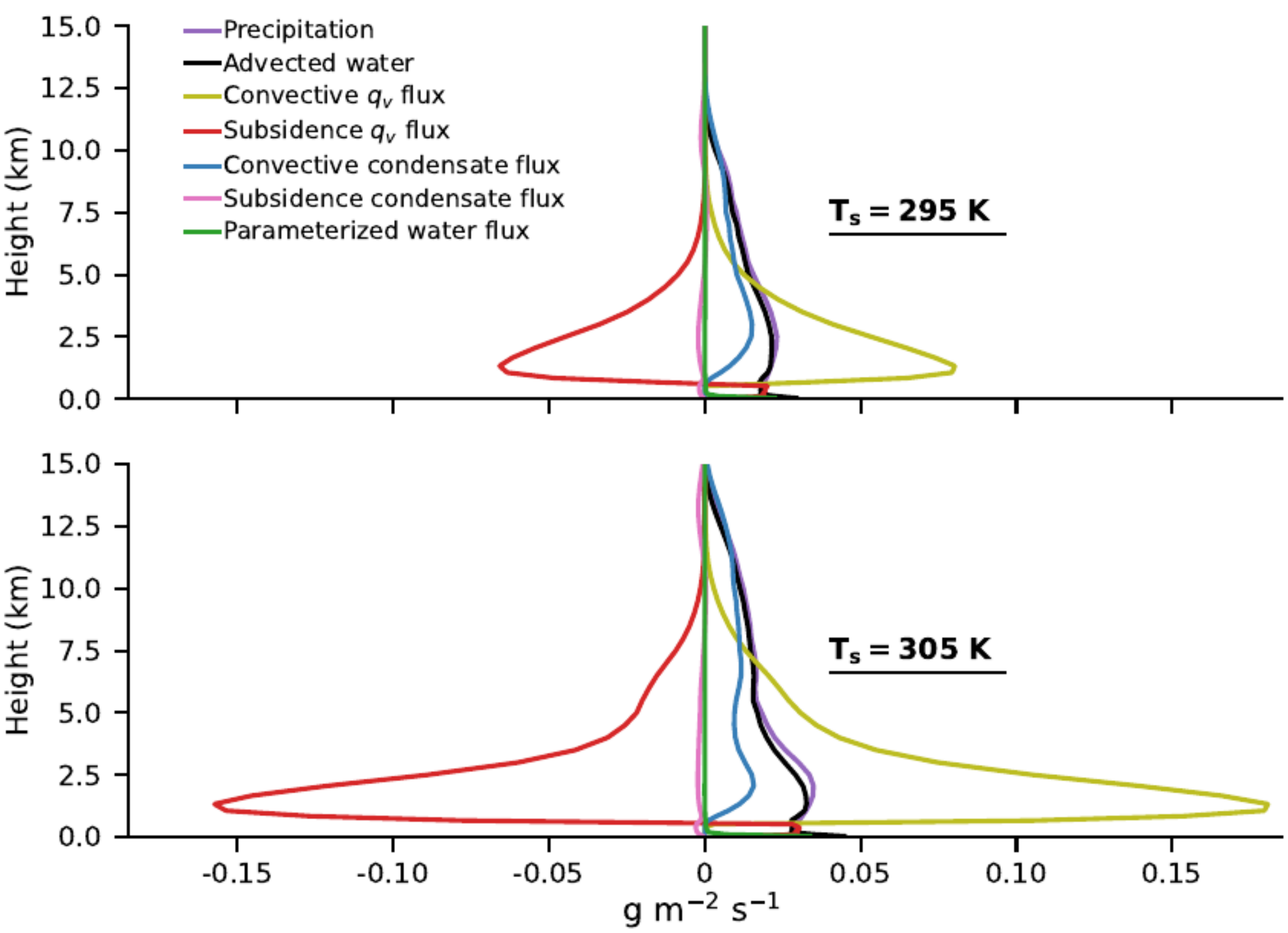

**Fig. S3 |** Water transport budget in SAM-P3ice. The full budget is shown for simulations with surface temperatures of 295 K (top row) and 305 K (bottom row). The method used to estimate convective and subsidence fluxes is described in Section 2.1.

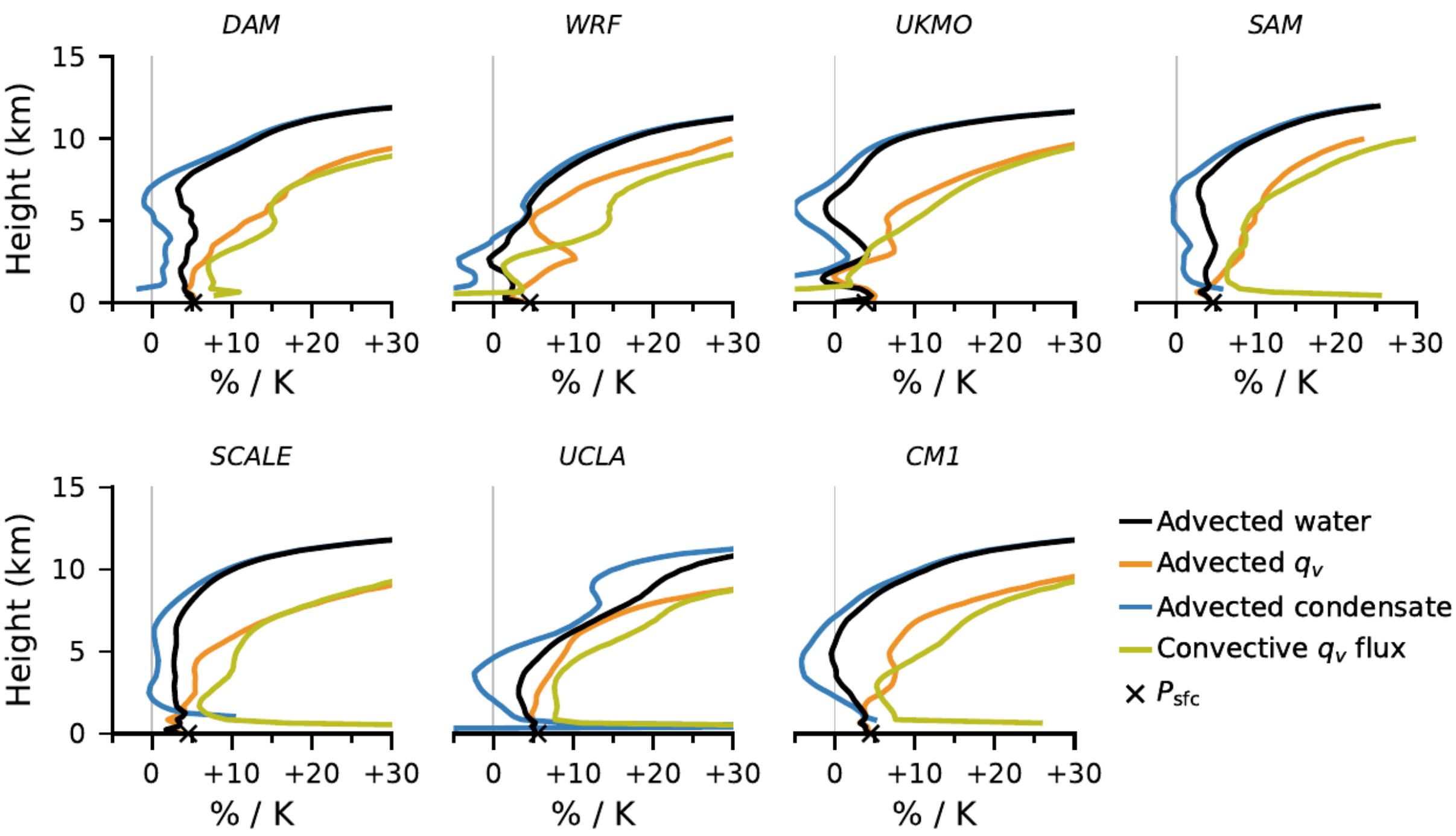

**Fig. S4 |** As in Fig. 4d but showing diagnostics separately across 7 RCEMIP models and SAM-P3ice.